\documentclass[conference]{IEEEtran}
\usepackage[utf8]{inputenc}
\usepackage[english]{babel}
\usepackage{cite}
\usepackage{amsmath,amssymb,amsfonts}
\usepackage{algorithmic}
\usepackage{graphicx}
\usepackage{textcomp}
\usepackage{xcolor}
\usepackage{mathrsfs} 

\usepackage{enumitem}
\newlist{legal}{enumerate}{10}
\setlist[legal]{label*=\arabic*.}
\usepackage{ifpdf}

\usepackage{amsthm}

\newtheoremstyle{mystyle1}
  {1pt}
  {1pt}
  {}
  {}
  {\bfseries}
  {.}
  { }
  {\thmname{#1}\thmnumber{ #2}\thmnote{ (#3)}}

\theoremstyle{mystyle1}

\newtheorem{rem}{Remark}
\newtheorem{exmp}{Example}

\newtheoremstyle{mystyle}
  {1pt}
  {1pt}
  {\itshape}
  {}
  {\bfseries}
  {.}
  { }
  {\thmname{#1}\thmnumber{ #2}\thmnote{ (#3)}}

 \theoremstyle{mystyle}

\newtheorem{thm}{Theorem}

\newtheorem{prop}{Proposition}

\usepackage{cite}

\ifCLASSINFOpdf
\else
\fi
\usepackage{amsmath}
\usepackage{array}
\begin{document}
%
\title{Independent User Partition Multicast Scheme for the Groupcast Index Coding Problem}

\author{\IEEEauthorblockN{Arman Sharififar, Neda Aboutorab}
\IEEEauthorblockA{\textit{School of Engineering and Information Technology} \\
\textit{University of New South Wales, Australia}\\
a.sharififar@student.unsw.edu.au\\ 
n.aboutorab@unsw.edu.au}
\and
\IEEEauthorblockN{Yucheng Liu, Parastoo Sadeghi}
\IEEEauthorblockA{\textit{Research School of Electrical, Energy and Materials Engineering} \\
\textit{Australian National University, Australia}\\
yucheng.liu@anu.edu.au\\
parastoo.sadeghi@anu.edu.au}
}
\maketitle


\begin{abstract}
The groupcast index coding (GIC) problem is a generalization of the index coding problem, where one packet can be demanded by multiple users. In this paper, we propose a new coding scheme called independent user partition multicast (IUPM) for the GIC problem. The novelty of this scheme compared to the user partition multicast (UPM) (Shanmugam \textit{et al.}, 2015) is in  removing redundancies  in  the UPM solution  by  eliminating  the  linearly dependent coded packets. We also prove that the UPM scheme subsumes the  packet partition multicast (PPM) scheme (Tehrani \textit{et al.}, 2012). Hence, the IUPM scheme is a generalization of both PPM and UPM schemes. Furthermore, inspired by jointly considering users and packets, we modify the approximation partition multicast (CAPM) scheme (Unal and Wagner, 2016)  to achieve a new polynomial-time algorithm for solving the general GIC problem. We characterize a class of GIC problems with $\frac{k(k-1)}{2}$ packets, for any integer $k\geq 2$, for which the IUPM scheme is optimal. We also prove that for this class, the broadcast rate of the proposed new heuristic algorithm is $k$, while the broadcast rate of the CAPM scheme is $\mathcal{O}(k^2)$.
\end{abstract}

\IEEEpeerreviewmaketitle

\section{Introduction}
This paper considers the problem of efficient broadcast in a system where a single server transmits a set of $m$ packets to a number of users via a noiseless broadcast channel. Each user requests one specific packet and may know some of the other packets a priori as its side information. This problem is known as the index coding problem and was first introduced by Birk and Kol \cite{Birk1998} in the context of satellite communications and further developed in \cite{5714242}. Since then, the index coding problem has attracted considerable attention from different research communities and various coding schemes have been proposed in the literature \cite{Shanmugam2013,Arbabjolfaei2014,Thapa2015,Agarwal2016,Thapa2017}. It has been shown that index coding problem offers a rich model that can be used to study several communication systems, such as distributed caching \cite{Maddah-Ali2014}, wireless network interference management \cite{Maleki2014}, and network coding \cite{ElRouayheb2010}.

In the unicast index coding (UIC) setting, each packet is demanded by exactly one user. However, in this paper, we consider a generalized index coding problem referred to as groupcast index coding (GIC), where more than one user is allowed to request the same packet. Such a scenario has been analyzed in \cite{Neely2013}, \cite{Alon2008}, where packet partition multicast (PPM) \cite{Tehrani2012} and user partition multicast (UPM) \cite{Shanmugam2014} schemes are proposed to solve the GIC problem. The PPM scheme proposed in \cite{Tehrani2012} works by partitioning the \textit{packets} into subsets and solving the problem for each subset independently. The authors in \cite{Tehrani2012} have shown that the PPM scheme is optimal for some families of the GIC problem. On the other hand, the UPM scheme proposed in \cite{Shanmugam2014} divides the problem into subproblems by partitioning the \textit{users} into subsets. Another coding scheme is proposed in \cite{Unal2016}, which is a polynomial-time heuristic algorithm, and due to its similarity with PPM, the authors refer to it as coded approximation partition multicast (CAPM). Reference \cite{Unal2016} further provides examples that show neither PPM nor CAPM outperforms the other scheme (see \cite[Examples 2 and 4]{Unal2016}). Moreover, a lower bound on the optimal broadcast rate of the GIC problem, namely the minimum lower bound (MLB), has been proposed in \cite{Unal2016}. 

For the UIC problem, there is a one-to-one mapping between packets and users. Thus, breaking a problem into a set of subproblems either by partitioning the packets or the users would be equivalent. However, for the GIC problem, since each packet can be demanded by multiple users, partitioning the users can lead to a better transmission rate.

In this paper, first we show that the PPM scheme is a special case of the UPM scheme. We also provide a class of GIC problems to illustrate how the UPM scheme can outperform the PPM scheme. In the UPM scheme, the users requesting the same packet can be placed into different subsets. Therefore, there is a chance that the users in one subset can use the coded packets of other subsets to recover their own packet. Motivated by this idea, we propose a new coding scheme, namely, independent user partition multicast (IUPM). Furthermore, unlike the CAPM algorithm where the initial subsets are formed by packet partitioning, we form the initial subsets by user partitioning to propose a new polynomial-time heuristic algorithm.
\subsection{Our Contributions}
The main contributions of this paper can be summarized as follows:
\begin{enumerate}
    \item We propose a new coding scheme, namely the IUPM scheme, which is an extension of the UPM scheme.
    \item We prove that the UPM scheme includes the PPM scheme as a special case. As a concrete performance comparison, we characterize a class of GIC problems, in which the number of packets is $m=\frac{k(k-1)}{2}$, each packet is requested by exactly two users and users' side information follow a certain pattern.  We refer to this class of GIC problems as $k$-Group 2-User GIC problems. We prove that for such class of problems the broadcast rate of the PPM scheme is lower bounded by $\frac{k(k-1)}{6}+1$, while the broadcast rate of the UPM scheme is $\mathcal{O}(k)$. We also prove that the IUPM scheme is optimal for this class of problems. 
    \item Based on user partitioning, we propose a new coding scheme. We show that the proposed scheme runs in polynomial time and for the aforementioned class of GIC problems with $k(k-1)/2$ packets, the broadcast rate of the new proposed heuristic algorithm is equal to $k$, while it is $\mathcal{O}(k^2)$ for the CAPM scheme.
\end{enumerate}
\subsection{Organization of the Paper}
The rest of this paper is organized as follows. In Section \ref{sec:2}, we review the system model and provide a brief overview of the PPM and UPM schemes. 
Section \ref{sec:3} provides a rigorous comparison between the UPM and PPM schemes in terms of their broadcast rate. In Section \ref{sec:4}, we propose our new coding scheme, the IUPM scheme, and prove that it is optimal for $k$-Group 2-User GIC problems. 
Section \ref{sec:5} proposes a new heuristic algorithm for the considered GIC problems based on user partitioning. Finally, Section \ref{sec:6} concludes the paper.
\section{System Model and Background}\label{sec:2}
\subsection{System Model}
In this paper, we consider a broadcast communication system in which a server transmits a set of $m$ packets $X=\{x_1,x_2,\dots,x_m\}, x_i\in\{0,1\}^{t_i}$ to a number of users via a noiseless broadcast channel. Let $U=U_1\cup U_2\cup\dots\cup U_m$ be the set of all users, where $U_i=\{u_i^{1},\dots,u_{i}^{|U_{i}|}\}$ is the set of users demanding the same packet $x_i$. Each user $u_{i}^{j}, i\in[m], j\in[|U_i|]$ (where $[a]:=\{1,2,\dots,a\}$) may know a subset of the packets $X_{i}^{j}:=\{x_{i^{\prime}} : i^{\prime}\in A_{i}^{j}\}, A_{i}^{j}\subseteq[m]\backslash \{i\}$, which is referred to as its side information set.  The main goal is to minimize the number of coded packets that the server should broadcast to the users so that each user is able to recover its desired packet.

We define $\mathcal{C}=(\phi,\{\psi_{i}^{j}\})$ as a $(t_1,\dots,t_m)$-bit index code, where $\phi$ and $\{\psi_{i}^{j}\}$ represent the encoder and decoder functions, respectively, and are defined as follows:

\begin{itemize}
    \item $\phi = \prod_{i\in[m]} \{0,1\}^{t_i}\rightarrow \{0,1\}^{r}$ is the encoder which maps the $m$ packets of $t_i$-bit, $i\in[m]$, to an $r$-bit output sequence.
    \item $\psi_{i}^{j}$ represents $|U|$ decoders, where for each user $u_{i}^{j}, i\in[m], j\in[|U_i|]$, the decoder $\psi_{i}^{j} = \{0,1\}^{r}\times \prod_{i^{\prime}\in A_{i}^{j}} \{0,1\}^{t_{i^{\prime}}}\rightarrow \{0,1\}^{t_i}$ maps the received $r$-bit sequence and the side information $X_{i}^{j}$ to a $t_i$-bit packet estimate of $x_i$.
\end{itemize}
If $t_i=t, \ \forall i\in[m]$, the broadcast rate of code $\mathcal{C}$ is $\beta(\mathcal{C})=\frac{r}{t}$. The optimal broadcast rate of the index coding problem is defined as

$$\beta=\inf_{t} \inf_{\mathcal{C}} \beta(\mathcal{C}).$$
Thus, the broadcast rate of each index code $\beta(\mathcal{C})$ provides an upper bound on $\beta$.

\subsection{Packet Partition Multicast (PPM) Scheme \cite{Tehrani2012}}
The PPM scheme starts with partitioning the packet set $X$ into $h$, $1 \leq h \leq m$, subsets $S_e\subseteq X, e\in[h] $ such that
\begin{equation} \label{eq:1}
   \left\{\begin{array}{lc}
        S_1 \cup ... \cup S_h=X,   \\ \\
        S_e \cap S_f=\emptyset, \ \ \text{if} \ e \neq f \ \text{for} \ e,f \in [h]. 
    \end{array}
    \right.
\end{equation}
Let $T_e=\{i\in[m] : x_i\in S_e\}$ be the set of indices of all packets inside $S_e$. Based on \eqref{eq:1}, we have
\begin{equation} \label{eq:2}
   \left\{\begin{array}{lc}
        T_1 \cup ... \cup T_h=[m],   \\ \\
        T_e \cap T_f=\emptyset, \ \ \text{if} \ e \neq f \ \text{for} \ e,f \in [h]. 
    \end{array}
    \right.
\end{equation}
We define $d_e$ as
\begin{equation} \label{eq:3}
    d_e=\min_{i\in T_e} \ \min_{j\in[|U_i|]} |A_{i}^{j}\cap T_e|.
\end{equation}
By transmitting $|T_e|-d_e$ coded packets using a $(|T_e|,|T_e|-d_e)$ maximum distance separable (MDS) code, all the users $u_{i}^{j}, i\in T_e, j\in[|U_i|]$ are able to recover their desired packet. Thus, for the set of subsets $T_e, e\in[h]$, the broadcast rate of the PPM scheme is equal to
\begin{equation} \label{eq:4}
   \beta_{PPM}(T_e,e\in [h])=\sum_{e=1}^{h} (|T_e|-d_e)=m-\sum_{e=1}^{h} d_e.
\end{equation} 
\begin{thm} [\mdseries Tehrani \textit{et al.} \cite{Tehrani2012}] \label{thm1}
   The optimal broadcast rate is upper bounded by $\beta_{PPM}$, which is the solution of the following optimization problem
    \begin{equation} \label{eq:5}
         \min_{T_e,e\in [h]} \beta_{PPM}(T_e,e\in [h]),
    \end{equation}
subject to the conditions in \eqref{eq:2}.
\end{thm} 
Note that finding the minimum of $\beta_{PPM}(T_e,e\in [h])$ is equivalent to solving the following optimization problem
\begin{equation} \label{eq:6}
\max_{T_e,e\in [h]} \sum_{e=1}^{h} d_e,
\end{equation}
which has been referred to as the sum-degree cover problem in \cite{Tehrani2012}.

Now we provide a simple GIC example to illustrate that by partitioning the users, a better performance than the PPM scheme can be achieved.
\begin{exmp} \label{exmp1}
    Consider a GIC problem with four packets and five users such that each packet is requested by one user except packet $x_1$, which is requested by two users. Therefore, the user set is $U=\{u_{1}^{1},u_{1}^{2},u_{2}^{1},u_{3}^{1},u_{4}^{1}\}$ and the side information of each user is as follows
    $$A_{1}^{1}=\{4\}, \ \ A_{1}^{2}=\{2,3\},\ \ A_{2}^{1}=\{1,3\},$$
    $$A_{3}^{1}=\{1,2\},\ \ A_{4}^{1}=\{1\}.$$
    Let us consider one of the packet partitions $S_1=\{x_2\},S_2=\{x_1,x_3,x_4\}$ with $h=2$. From \eqref{eq:3}, $d_1=0$, $d_2=1$, and the broadcast rate for this partition will be 3. One can verify that the broadcast rate for other partitions cannot be better than this partition. Therefore, $\beta_{PPM}=3$. However, if instead of partitioning the packets, we partition the users into two subsets $\{u_{1}^{1},u_{4}^{1}\}$ and $\{u_{1}^{2},u_{2}^{1},u_{3}^{1}\}$, then by sending $p_1\oplus p_4$  and $p_1\oplus p_2 \oplus p_3$ the users in $\{u_{1}^{1},u_{4}^{1}\}$ and $\{u_{1}^{2},u_{2}^{1},u_{3}^{1}\}$ are able to decode their desired packet, respectively. Therefore, for this user partitioning the broadcast rate is 2. The MLB in \cite{Unal2016} for this example is also 2. Therefore, this user partitioning is optimal.
\end{exmp} 
\subsection{User Partition Multicast (UPM) Scheme \cite{Shanmugam2014}}
The UPM scheme starts with partitioning the user set $U$ into $h$, $1 \leq h \leq m$, subsets $W_e \subseteq U, e\in[h]$ such that
\begin{equation} \label{eq:7}
    \left\{\begin{array}{lc}
         W_1\cup\dots\cup W_h=U, \ \ \ \ \ \ \ \  \\ \\
         W_e \cap W_f=\emptyset, \ \ \text{if} \ e \neq f \ \text{for} \ e,f \in [h].
    \end{array}\right.
\end{equation}
Let $Y_e=\{i\in[m] : u_{i}^{j}\in W_e \ \text{for some}\ j\in[|U_i|]\}$ and $Z_{e,i}=\{j\in[|U_i|] : u_{i}^{j}\in W_e\}$. Based on \eqref{eq:7}, we will have
\begin{equation} \label{eq:8}
    \left\{\begin{array}{lcc} 
         Y_1\cup\dots\cup Y_h=[m], \ \ \ \ \ \ \\ \\
         Z_{1,i}\cup\dots\cup Z_{h,i}=[|U_i|], \ \ \ \ \ i\in [m], \\ \\
         Z_{e,i} \cap Z_{f,i}=\emptyset, \ \ \ \text{if} \ e \neq f \  \ \ \text{for} \ e,f \in [h], \forall i\in [m].
    \end{array}\right.
\end{equation}
We define $c_e$ as
\begin{equation} \label{eq:9}
    c_e=\min_{i\in Y_e} \ \min_{j\in Z_{e,i}} |A_{i}^{j}\cap Y_e|.
\end{equation}
By sending the $|Y_e|-c_e$ coded packets using a $(|Y_e|, |Y_e|-c_e)$ MDS code, all the users $u_{i}^{j}\in W_e$ can decode their desired packet. Thus,  for the set of subsets $Y_e,Z_{e,i}, e\in[h],i\in [m]$, the broadcast rate  of the UPM scheme is equal to
\begin{equation} \label{eq:10}
    \beta_{UPM}(Y_e,Z_{e,i}, e\in[h],i\in [m])=\sum_{e=1}^{h} (|Y_e|-c_e).
\end{equation}
\begin{thm} [\mdseries Shanmugam \textit{et al.}\cite{Shanmugam2014}] \label{thm2}
The optimal broadcast rate is upper bounded by $\beta_{UPM}$, which is the solution of the following optimization problem
\begin{equation} \label{eq:11}
     \min_{Y_e,Z_{e,i}, e\in[h],i\in [m]} \beta_{UPM}(Y_e,Z_{e,i}, e\in[h],i\in [m]),
\end{equation}
subject to the conditions in \eqref{eq:8}.
\end{thm} 
\section{Comparison Between PPM and UPM schemes}\label{sec:3} 
This section provides a rigorous comparison of the PPM and UPM schemes, and then investigates their performance for the class of GIC problems considered in this paper.
\begin{prop} \label{prop1}
     The PPM scheme is a special case of the UPM scheme.
\end{prop}
\begin{IEEEproof} In the UPM scheme, $Z_{e,i}\subseteq[|U_i|]$. If we only consider $Z_{e,i}=[|U_i|]$ for some $e \in [h]$, according to the third condition in \eqref{eq:8}, we have
\begin{equation}\label{eq:prop1}
Y_e \cap Y_f=\emptyset, \ \ \text{if} \ e \neq f \ \text{for} \ e,f \in [h].
\end{equation}
Therefore, \eqref{eq:9} becomes identical to \eqref{eq:3}, the conditions in \eqref{eq:8} become identical to the conditions in \eqref{eq:2}, and therefore, the optimization problem in \eqref{eq:11} will become equal to the optimization problem in \eqref{eq:5}. This shows that for any $Z_{e,i}=[|U_i|]$, the UPM scheme subsumes all searches over packet partitions in the PPM scheme to minimize the broadcast rate.
\end{IEEEproof}
\begin{rem}
    Reference \cite{Tehrani2012} shows that the PPM scheme is optimal for three families of index coding problems: 1) directed cycles problem, 
    2) complete side information problem, and 
    3) directed acyclic problem. 
    Given that the PPM scheme is a special case of the UPM scheme (based on Proposition \ref{prop1}), the latter will also be optimal for these three cases.
\end{rem}  


\begin{rem}
       Reference \cite{Tehrani2012} proves that overlapping packets between two subsets in \eqref{eq:1} can decrease $\beta_{PPM}$. However, in the UPM scheme, users that want the same packet may be placed into different subsets. Therefore, the same requested packet needs to be transmitted in both subproblems. 
\end{rem} 


Now we characterize a class of GIC problems with $m=\frac{k(k-1)}{2}$ packets for any integer $k\geq 2$, such that each packet is requested by two users  $U_i=\{u_{i}^{1},u_{i}^{2}\}, \forall i\in[m]$. There is also a certain structure to users' side information set as follows. First, for $l \in [k]$, define two sets
\begin{equation} \label{eq:13}
\left\{\begin{array}{lc}
       I_{l}^{1} &=\bigcup_{a=1}^{k-l} \{{(l-1)k+(a-\frac{l(l-1)}{2})}\}, \ \ l\neq k  \ \ \ \\ \\
       I_{l}^{2} &=\bigcup_{a=1}^{l-1} \{{(a-1)k+(l-\frac{a(a+1)}{2})}\}, \ \ l\neq 1. \ \ \
\end{array} \right. 
\end{equation}
In addition, we have $I_{k}^{1}=I_{1}^{2} =\emptyset$ and $I_{l} = I_{l}^{1} \cup I_{l}^{2}$. 
Each user $u_{i}^{j}$ has the following structure to its side information
\begin{equation}
    A_{i}^{j}=I_{l}\backslash \{i\}, \ \ \ \text{if} \ i\in I_{l}^{j}.\label{eq:111}
\end{equation}
We refer to this class of problems with the aforementioned special side information structure as the $k$-Group 2-User GIC problems and denote it by $(k,2)$-GIC problems. We can utilise the structure of user side information to divide the users into $k$ groups $G_l, \ l\in[k]$ as $G_{l}= G_{l}^{1} \cup G_{l}^{2}$ where
\begin{equation}
\left\{\begin{array}{lc} \label{eq:12}
       G_{l}^{1} &=\{u_{i}^{1} : i\in I_{l}^{1}\}, \ \ \ \ \ l\neq k,   \\ \\
       G_{l}^{2} &=\{u_{i}^{2} : i\in I_{l}^{2}\}, \ \ \ \ \ l\neq 1.
\end{array} \right.  \ \ \ \ \ \ \ \ \ \ \ \ \ \ \ \ \
\end{equation}
 In addition, we have $G_{k}^{1} = G_{1}^{2} =\emptyset$. In words, each user in each group has the  packets  requested  by  the  users  in  the  same  group  as its side information set.
\begin{exmp} \label{exmp2}
   If we select $k=6$, then we have $m=\frac{k(k-1)}{2}=15$ packets, $|U|=k(k-1)=30$ users, and the six user groups are as follows
\begin{equation}
\left\{\begin{array}{lcc} \nonumber
    G_1=\{u_1^{1},u_2^{1},u_3^{1},u_4^{1},u_5^{1}\},\ \ \ G_2=\{u_1^{2},u_6^{1},u_7^{1},u_8^{1},u_9^{1}\},\ \ \ \ \ \\ \\
    G_3=\{u_2^{2},u_6^{2},u_{10}^{1},u_{11}^{1},u_{12}^{1}\},G_4=\{u_3^{2},u_7^{2},u_{10}^{2},u_{13}^{1},u_{14}^{1}\}, \\ \\
    G_5=\{u_4^{2},u_8^{2},u_{11}^{2},u_{13}^{2},u_{15}^{1}\},G_6=\{u_5^{2},u_9^{2},u_{12}^{2},u_{14}^{2},u_{15}^{2}\}.
    \end{array}\right. 
\end{equation}
\end{exmp} 

Now we investigate the performance of the PPM and UPM schemes for the $(k,2)$-GIC problems.
\subsection{Performance of the PPM Scheme for $(k,2)$-GIC Problems}
\begin{prop} \label{prop2}
   The broadcast rate of the PPM scheme for the $(k,2)$-GIC problems is lower bounded as
\begin{equation}
\beta_{PPM} \geq \ \frac{k(k-1)}{6}+1.
\end{equation}
\end{prop}
\begin{IEEEproof}
  First we express the packet set $X$ and its subsets $S_{e},e\in[h]$ based on $I_l,l\in[k]$ as $X=\{x_i : i\in I_{1}\cup\dots\cup I_{k}\}$, and $S_e=\{x_i : i\in I_{1,e}\cup\dots\cup I_{k,e}\}$, where $I_{l,e}=I_{l}\cap T_e$.\\
For satisfying the conditions in \eqref{eq:2}, we must have
\begin{equation} \label{eq:14}
   \left\{\begin{array}{lc}
        I_{l}=I_{l,1} \cup\dots\cup I_{l,h},  \\ \\
        I_{l,e} \cap I_{l,f}=\emptyset, \ \ \text{if} \ e \neq f \ \text{for} \ e,f \in[h].
    \end{array}
    \right.
\end{equation}
According to \eqref{eq:111}, we have
\begin{equation} 
    A_{i}^{j}\cap T_e=(I_l\backslash\{i\})\cap T_e=I_{l,e}\backslash\{i\},
\end{equation}
where $I_{l,e}\neq \emptyset$. Thus, the minimum size of the local side information of each subproblem can be obtained as
\begin{equation} 
     d_e=\min_{i\in T_e} \ \min_{j\in\{1,2\}} |A_{i}^{j}\cap T_e| 
     =\min_{|I_{l,e}|\neq0} (|I_{l,e}|-1).
\end{equation}
Therefore, the optimization problem in \eqref{eq:6} can be expressed as
\begin{equation} 
   \max_{I_{l,e},e\in[h]} \sum_{e=1}^{h} d_e=\max_{I_{l,e},e\in[h]} \sum_{e=1}^{h} \min_{|I_{l,e}|\neq0} (|I_{l,e}|-1).\\
\end{equation}
Let $N_{e}=\{l\in [k] : |I_{l,e}|\neq 0\},\forall e\in [h]$. Every subset must have at least one packet and according to \eqref{eq:12} this packet is placed into two groups. So, $|N_{e}|\geq 2$. Moreover, if subset $S_e$ has only one packet, then $|I_{l,1}|=1$, which cannot increase $d_e$ since $|I_{l,1}|-1=0$. Thus, each subset should have at least two packets and according to \eqref{eq:13}, it can be verified that every two packets are placed into three or four groups. So, $|N_{e}|\geq 3$.
 Thus, we have
\begin{align}
    \max_{I_{l,e},e\in[h]} \sum_{e=1}^{h} \min_{l\in N_e} (|I_{l,e}|-1)& \leq \max_{I_{l,e},e\in[h]} \sum_{e=1}^{h} \sum_{l\in N_e} \frac{|I_{l,e}|-1}{|N_e|} \nonumber \\
    & \leq \max_{I_{l,e},e\in[h]} \sum_{e=1}^{h} \sum_{l\in N_e} \frac{|I_{l,e}|-1}{3} \nonumber \\
    & = \ \ \sum_{l=1}^{k} (\frac{I_l}{3})- \frac{|N_e|h}{3} \label{eq:15}\\
    & \leq \ \ \frac{k(k-1)}{3}-h \nonumber\\
    & \leq \ \ \ \frac{k(k-1)}{3}-1 \label{eq:16},
\end{align}
where \eqref{eq:15} is due to \eqref{eq:14} and \eqref{eq:16} is due to $h\geq 1$.\\
Therefore, for the broadcast rate of the PPM scheme, we have
\begin{equation} 
  \beta_{PPM}=\frac{k(k-1)}{2}-\max_{I_{l,e},e\in[h]} \sum_{e=1}^{h} d_e \ \geq \ \frac{k(k-1)}{6}+1.
\end{equation}
\end{IEEEproof}
\subsection{Performance of the UPM Scheme for $(k,2)$-GIC Problems}
\begin{prop} \label{prop3}
   The broadcast rate of the UPM scheme for the $(k,2)$-GIC problems is upper bounded by k.
\end{prop} 
\begin{IEEEproof} 
It can be verified that the groups $G_l, \ l\in [k],$ in \eqref{eq:12} satisfy the two conditions in \eqref{eq:7} as follows
\begin{equation}
    \left\{\begin{array}{lc}
         U=G_1\cup\dots\cup G_{k}, \ \ \ \ \ \ \ \ \ \\ \\
         G_e \cap G_f=\emptyset, \ \ \text{if} \ e \neq f \ \text{for} \ e,f \in[k].
    \end{array}\right.
\end{equation}
Therefore, we have $k$ subsets, each including $k-1$ users demanding different packets. Then, for each subset $S_e, e\in[k]$ we have
\begin{equation}
|Y_e|=k-1, \ \ \ \ \ \ \ c_e=k-2.
\end{equation}
According to \eqref{eq:10}, the broadcast rate of the UPM is
\begin{equation}
\beta_{UPM}\leq \sum_{e=1}^{k} (|Y_e|-c_e)=k.
\end{equation}
In fact, by sending $\oplus_{i\in I_l} x_i$, the users in $G_l$ are able to recover their packets.
\end{IEEEproof}

\section{The Proposed Independent User Partition Multicast (IUPM) Scheme}\label{sec:4}

\subsection{The Idea Behind the IUPM Scheme}
By performing an exhaustive search over $k=4,5,6,7$, it can be verified that the broadcast rate of the UPM scheme for the class of $(k,2)$-GIC problems is equal to $k$. For instance, in Example \ref{exmp2}, the UPM scheme sends six $w_l=\oplus_{i\in I_l} x_i, l\in[6]$ coded transmissions. However, it can be shown that one of the coded packets can be simply obtained by XORing the others (for example, $w_6=\oplus_{l\in[5]} w_l$). Thus, the broadcast rate of the UPM scheme can be further improved by checking whether the coded packets for user subsets are linearly dependent or not. The saving can be significant as shown through the following example. 
\begin{exmp}
    Consider a GIC problem with $m=20$ packets such that each packet is demanded by 3 users ($|U_i|=3, i\in[20]$). We divide the users into 15 groups such that each user has the packets requested by the users in the same group as its side information set. It means
\begin{equation} \nonumber
\left\{\begin{array}{lccccccc} 
    G_1=\{u_1^{1},u_2^{1},u_{3}^{1},u_{4}^{1}\},\ \ \ \ \ G_2=\{u_1^{2},u_5^{1},u_{6}^{1},u_{7}^{1}\},\ \ \ \ \ \ \ \\ \\
    G_3=\{u_2^{2},u_5^{2},u_{8}^{1},u_{9}^{1}\},\ \ \ \ \ G_4=\{u_3^{2},u_6^{2},u_{8}^{2},u_{10}^{1}\}, \ \ \ \ \ \ \\ \\
    G_5=\{u_4^{2},u_{7}^{2},u_{9}^{2},u_{10}^{2}\},\ \ \ \ G_6=\{u_1^{3},u_{11}^{1},u_{12}^{1},u_{13}^{1}\}, \ \ \ \ \\ \\
    G_7=\{u_2^{3},u_{11}^{2},u_{14}^{1},u_{15}^{1}\},\ \  G_8=\{u_3^{3},u_{12}^{2},u_{14}^{2},u_{16}^{1}\}, \ \ \ \\ \\
    G_9=\{u_{4}^{3},u_{13}^{2},u_{15}^{2},u_{16}^{2}\},\ \ \ G_{10}=\{u_5^{3},u_{11}^{3},u_{17}^{1},u_{18}^{1}\}, \ \\ \\
    G_{11}=\{u_{6}^{3},u_{12}^{3},u_{17}^{2},u_{19}^{1}\},\ \ G_{12}=\{u_{7}^{3},u_{13}^{3},u_{18}^{2},u_{19}^{2}\}, \ \ \\ \\
    G_{13}=\{u_{8}^{3},u_{14}^{3},u_{17}^{3},u_{20}^{1}\},\ \ \ G_{14}=\{u_{9}^{3},u_{15}^{3},u_{18}^{3},u_{20}^{2}\},\ \\ \\
    G_{15}=\{u_{10}^{3},u_{16}^{3},u_{19}^{3},u_{20}^{3}\}. \ \ \ \ \ \ \ \ \ \ \ \ \ \ \ \ \ \ \ \ \ \ \ \ \ \ \ \ \ \ \ \ \ \
    \end{array}\right. 
\end{equation}
For this example, it can be verified (through an exhaustive search) that the optimal user subsets are $W_e=G_e, e\in[15]$ showing that the broadcast rate of the the UPM scheme is 15. However, it can be seen that the users in $G_5,G_9,G_{12},G_{14},G_{15}$ are able to decode their desired packets by the coded transmissions of the other groups. It means
\begin{equation} \nonumber
    \begin{array}{lll}
        w_5\ =w_1\oplus w_2\oplus w_3\oplus w_4,\\
        w_9\ =w_1\oplus w_6\oplus w_7\oplus w_8,\\
        w_{12}=w_2\oplus w_6\oplus w_{10}\oplus w_{11},\\ 
        w_{14}=w_3\oplus w_7\oplus w_{10}\oplus w_{13},\\
        w_{15}=w_4\oplus w_8\oplus w_{11}\oplus w_{13},
    \end{array}
\end{equation}
where $w_e=\oplus_{i\in Y_e} x_i$.
This shows we can save five transmissions, matching the MLB \cite{Unal2016}, thus achieving the optimal broadcast rate of $\beta = 10$.
\end{exmp} 

\subsection{Formal Description of the IUPM Scheme}
Let $b_e=|Y_e|-c_e$, $g_e=\sum_{f\in[e]} b_f$ and $g_{0}=1$. Recall that for subset $W_e$, the UPM scheme sends $b_e$ linear equations as $\sum_{i\in Y_e} \alpha_{i,(g_{e-1})+1}x_i,\dots,\sum_{i\in Y_e} \alpha_{i,g_{e}}x_i$ where $\alpha_{i,j}$'s are taken from some finite field $\mathbb{F}$ and $\alpha_{i,j}=0, \forall i\notin Y_e$. For the IUPM scheme, however, we remove the equations that are linearly dependent on each others. Therefore, we construct a matrix $\mathbf{N}$ with elements represented by $n_{i,j}$, where $n_{i,j}=\alpha_{i,j}$. Having defined $\mathbf{N}$, for the set of subsets $Y_e,Z_{e,i}, e\in[h],i\in [m]$, the broadcast rate of the IUPM scheme is equal to
\begin{equation}
    \beta_{IUPM}(Y_e,Z_{e,i}, e\in[h],i\in [m])=\min_{\alpha_{i,j},i\in[m],j\in[g_h]} \mathrm{rank} \ \mathbf{N}
\end{equation}
\begin{thm} \label{thm3}
  For a given GIC problem, the optimal broadcast rate is upper bounded by $\beta_{IUPM}$, which is the solution to the following optimization problem:
\begin{equation}
     \min_{Y_e,Z_{e,i}, e\in[h],i\in [m]} \beta_{IUPM}(Y_e,Z_{e,i}, e\in[h],i\in [m]).
\end{equation}
\end{thm}
\begin{rem}
   It is not possible to improve the broadcast rate of the PPM scheme by choosing just those transmissions that are linearly independent. The reason is that in the PPM scheme, based on \eqref{eq:2}, $T_e\cap T_f=\emptyset, \ \ \text{if} \ e \neq f \ \text{for} \ e,f \in [h]$. In words, each packet is transmitted in just one subproblem. Therefore, the columns of $\mathbf{N}$ are all linearly independent. 
   \end{rem}
 \begin{rem}
    For the unicast index coding problem, the PPM, UPM, and IUPM schemes are equivalent, as they all can be reduced to the partial clique covering scheme \cite{Birk1998}. In fact, it is the GIC setup (i.e., where the users that are partitioned into different subproblems may require the same packet) that provides the opportunity for the users in one subproblem to leverage coded packets from some other subproblems and decode their required packets.
\end{rem}

\begin{prop} \label{prop4}
     IUPM scheme is optimal for the $(k,2)$-GIC problems.
\end{prop} 
\begin{IEEEproof}
As mentioned earlier, for the user groups in \eqref{eq:12}, sending $\oplus_{i\in I_l} x_i$ satisfies the users in $G_l$. Now, we prove that $\oplus_{l\in [k-1]}(\oplus_{i\in I_l} x_i)=\oplus_{i\in I_k} x_i$, which means that the users in group $G_k$ can decode their packet by XORing the coded transmissions of the other groups. From \eqref{eq:13}, it can be easily shown that $p_{l_1}^{1}(l_2-l_1)=p_{l_2}^{2}(l_1)$ for $l_2>l_1$, where $l_1,l_2\in[k]$, and
\begin{equation}
    \left\{\begin{array}{lc}
             p_{l}^{1}(a)&={(l-1)k+(a-\frac{l(l-1)}{2})},\ \ \\ \\
             p_{l}^{2}(a)&={(a-1)k+(l-\frac{a(a+1)}{2})},
    \end{array} \right. 
\end{equation}
which means $I_{l_1}\cap I_{l_2}=\{p_{l_2}^{2}(l_1)\}$. Since each packet is transmitted twice (for two users), XORing the coded packets $\oplus_{i\in I_l} x_i$ for $l\in[k-1]$ gives us the XOR of the packets in
\begin{equation} \nonumber
    \cup_{i\in[k-1]} I_l\backslash\cup_{\substack{l_1,l_2\in[k-1]\\l_2> l_1\ \ \ \ \ \ \ }}(I_{l_1}\cap I_{l_2})=\cup_{l_1\in[k-1]} \{p_{k}^{2} (l_1)\}=I_k,
    \end{equation}
and as a result $\oplus_{l\in [k-1]}(\oplus_{i\in I_l} x_i)=\oplus_{i\in I_k} x_i$. Therefore, $\beta_{IUPM}=k-1$, which achieves the MLB \cite{Unal2016}.
\end{IEEEproof}
\section{The Proposed Heuristic Algorithm for the GIC Problems Based on User Partitioning}\label{sec:5}
The proposed heuristic algorithm is a new polynomial-time heuristic algorithm, consisting of three steps, which is based on user partitioning. It can be considered as a modified version of the CAPM scheme \cite{Unal2016}, where instead of packet partitioning, the algorithm works based on partitioning of the users to improve the broadcast rate as explained in the remainder of this section. The proposed algorithm is based on the following three steps:
\textit{Step 1.} The algorithm starts with initializing the user partitions, such that each user $u_{i}^{j}$ is assigned an initial subset based on both its side information set and the users which have packet $x_i$ in their side information set\footnote{It is worth mentioning that this is unlike the CAPM algorithm which starts with assigning a subset to each packet $x_i$ (instead of each user $u_{i}^{j}$).}.\\
Let $U_{i}^{\prime}$ be the set of users which have packet $x_i$ in their side information set. For describing how the proposed heuristic algorithm and the CAPM scheme work differently in this step, we denote each user $u_{i}^{j}$ as a tuple $(i,j)$. Assume $V_i$ and $V_{i}^{\prime}$ denote the indices of users in $U_i$ and $U_{i}^{\prime}$, respectively, as
\begin{equation}
    V_{i}=\{(i,j) : j\in[|U_i|]\}, V_{i}^{\prime}=\{(i_1,j) : u_{i_1}^{j}\in U_{i}^{\prime}\}.
\end{equation}
In the proposed heuristic algorithm, we consider an initial subset for each user. First, we use $B_{i}^{j}$ to represent the side information $A_{i}^{j}$ as a tuple, as
\begin{equation}
B_{i}^{j}=\cup_{i^{\prime}\in A_{i}^{j}} \{(i^{\prime},j^{\prime}) : j^{\prime}\in[|U_{i^{\prime}}|]\}.
\end{equation}
Then, we assign to user $u_{i}^{j}$ a subset as follows
\begin{equation} \label{eq:20}
    S_{V_{i,j}^{\star}}=\{x_i\}, \text{ where }  V_{i,j}^{\star}=(i,j)\cup (B_{i}^{j} \cap V_{i}^{\prime}).
\end{equation}
If $v_{i,j}^{\star}=|V_{i,j}^{\star}|$, the subset $S_{V_{i,j}^{\star}}$ is called a level-$v_{i,j}^{\star}$ packet.

The complexity of \textit{Step 1} is at most $\mathcal{O}(m.|U|^2)$.\\
\textit{Step 2.}
The second step begins by considering the subsets with the lowest level packet and checking the entropy value $H(S_{V_{i,j}^{\star}}|A_{i}^{j})$ for all $(i,j)\in V_{i,j}^{\star}$. If the entropy is zero for some $(i,j)\in V_{i,j}^{\star}$, then the packets in $S_{V_{i,j}^{\star}}$ are moved to the next higher level. If there is already one or more subsets in the next higher level, the packets are moved to one of them arbitrarily. Otherwise, the subset with a higher level can be obtained by adding the lowest indices to $V_{i,j}^{\star}$, which are not already in $V_{i,j}^{\star}$. For the order of indices, we have $(i,j)<(i',j')$, if
\begin{equation} \label{eq:18}
    \begin{cases} 
    i<i' \ \ \  \text{if} \ i\neq i',\\
    j<j' \ \ \ \text{if} \ i= i'.
    \end{cases}
\end{equation}
This process is repeated until this subset merges with another subset. It means that as a result of adding the lowest index to the index of a subset, the final index would become the same for some subsets and therefore, their assigned packets will be placed into the subset with the same index. Then, the step is repeated by checking the entropy value until it becomes positive and equal for all $(i,j)\in V_{i,j}^{\star}$, and then we go to the next step. The complexity of \textit{Step 2} is at most $\mathcal{O}(m^2.|U|^3)$.\\
\textit{Step 3.} In the third step, if there are some packets in the same subset in Step 1, they are replaced with their XOR. Assume that $S_{K_1},...,S_{K_q}$ are the final subsets at the end of Step 3. Then, the broadcast rate of the proposed heuristic algorithm will be
\begin{equation} \label{eq:19}
    \beta_{heuristic}=\sum_{p\in [q]} \max_{(i,j) \in K_p} H(S_{K_p} \mid A_{i}^{j}).
\end{equation}
The complexity of \textit{Step 3} is at most $\mathcal{O}(m.|U|^3)$.

\subsection{The Broadcast Rate of the Proposed Heuristic Algorithm for the $(k,2)$-GIC Problems}
This subsection establishes the the broadcast rate of the proposed heuristic algorithm, as well as that of the CAPM scheme for the class of $(k,2)$-GIC problems. We prove that the broadcast rate of the proposed heuristic algorithm is $k$, while the broadcast rate of the CAPM scheme is $\frac{k(k-3)}{2}+2$, which is $\mathcal{O}(k^2)$.
\begin{prop} \label{prop5}
     The broadcast rate of the proposed heuristic scheme for the class of $(k,2)$-GIC problems is $k$.
\end{prop}
\begin{IEEEproof}
For this class of the problems, the three steps of the heuristic algorithm are explained below.\\
\textit{Step 1}: From \eqref{eq:20}, it can be verified that the users of each group are placed in the same subset
\begin{equation} 
    S_{V_l}=\{x_i : i\in I_l\} \ \ l\in[k],
\end{equation}
which implies that the packets requested by users in each group are included in the same subset.\\
\textit{Step 2}: It can be verified that
\begin{equation} 
    H(S_{V_l}|A_{i}^{j})=1, \ \ \ \forall (i,j)\in V_l.
\end{equation}
Since the entropy value for all users in the same partition is equal, we proceed to the next step.\\
\textit{Step 3}: Note that at Step 1, packets $x_i, i\in I_l$ are in the same subset. Therefore, they can be XORed together and as a result, the broadcast rate will be
\begin{equation} 
    \beta_{heuristic}=\sum_{l\in [k]} \max_{(i,j)\in V_l} H(S_{V_l}|A_{i}^{j})=k.
\end{equation}
\end{IEEEproof}
\begin{prop} \label{prop6}
    The broadcast rate of the CAPM scheme for the class of $(k,2)$-GIC problems is $\frac{k(k-3)}{2}+2$.
\end{prop}
\begin{IEEEproof}
Suppose that $V_{l}$ is the set of indices of users in $G_{l}$. It means
\begin{equation}
\left\{\begin{array}{ccc}
       V_{l}^{1}=\{(i,1) : i\in I_{l}^{1}\}, \ \ \ \ \ l\neq k, \ \ \ \ \ \ \ \ V_{k}^{1}=\emptyset,   \\ \\
       V_{l}^{2}=\{(i,2) : i\in I_{l}^{2}\}, \ \ \ \ \ l\neq 1, \ \ \ \ \ \ \ \ V_{1}^{2}=\emptyset, \\ \\
       V_{l} = V_{l}^{1} \cup V_{l}^{2}. \ \ \ \ \ \ \ \ \ \ \ \ \ \ \ \ \ \ \ \ \ \ \ \ \ \ \ \ \ \ \ \ \ \ \ \ \ \ \ \
\end{array} \right.
\end{equation} 
And also $V_{l}^{\prime}=\{(i,2) : i\in I_{l}^{1}\},l\neq 1,V_{1}^{\prime}=\emptyset$. For any arbitrary set like $R$, we define $R_{l_1,\dots,l_{k^{\prime}}}^{}:=\cup_{b=1}^{k^{\prime}} R_{l_b}.$
We now follow the three steps of the CAPM scheme for the class of $(k,2)$-GIC problems (for more details on the CAMP algorithm, see \cite{Unal2016}).\\
\textit{Step 1}: Unlike the proposed algorithm, in the CAPM scheme, 
the packets are placed into an initial partition as follows
$$S_{V_{l_1,l_2}}=\{x_i : i\in I_{l_1}\cap I_{l_2}\}, \ \ \forall l_1\neq l_2\in[k].$$
Then the rest of the procedure is similar to the proposed heuristic algorithm.\\
\textit{Step 2}: Since each packet $x_i$ is demanded by two users $u_{i}^{1},u_{i}^{2}$, for all $i\in[m]$, we have
\begin{equation} \nonumber
    H(S_{V_{l_1,l_2}}\mid A_{i}^{j})=0, \ \ j\in \{1,2\}.
\end{equation}
Therefore, we need to move all the packets to the next higher level by adding to each subset the lowest indices that are not already in $V_{l_1,l_2}$ until two or more of them are merged with each other. According to \eqref{eq:18}, for the order of indices we have
\begin{equation} 
    V_{1}^{1} \rightarrow V_{1}^{\prime} \rightarrow V_{2}^{1} \rightarrow V_{2}^{\prime} \rightarrow \dots \rightarrow V_{k}^{1} \rightarrow V_{k}^{\prime}.
\end{equation}
After adding the lowest indices to $V_{l_1,l_2}$, it can be seen that $S_{V_{1,2}}$,$S_{V_{1,3}}$, and $S_{V_{2,3}}$ are the first subsets that merge with each other as follows
$$S_{V_{1,2,3}^{1}\cup V_{1,2,3}^{\prime}}=\{x_i : i\in \cup_{\substack{l_1,l_2\in[3]\\l_1\neq l_2\ \ \ }} (I_{l_1}\cap I_{l_2})\}.$$
Since $H(S_{V_{1,2,3}^{1}\cup V_{1,2,3}^{\prime}} \mid A_{i}^{2})=0, \forall i\in I_{1,2}^{1}\backslash I_{1,2,3}^{2}$, we have to repeat the process again. It can be observed that this process continues as follows
\begin{equation} \nonumber
    S_{V_{1,\dots,k-1}^{1}\cup V_{1,\dots,k-1}^{\prime}}=\{x_i : i\in \cup_{\substack{l_1,l_2\in[k-1]\\l_1\neq l_2\ \ \ \ \ \ }} (I_{l_1}\cap I_{l_2})\},
\end{equation}
\begin{equation} \nonumber
    H(S_{V_{1,\dots,k-1}^{1}\cup V_{1,\dots,k-1}^{\prime}} \mid A_{i}^{2})=0, \ \ \ \forall i \in I_{1,...,k-2}^{1}\setminus I_{1,...,k-1}^{2}.
\end{equation}
Therefore, we move all the packets to the highest-level subset which includes the indices of all the users. At the end of this step, we will have
\begin{equation} \nonumber
     S_{V_{1,\dots,k}^{1}\cup V_{1,\dots,k}^{\prime}}=\{x_i : i\in \cup_{\substack{l_1,l_2\in[k]\\l_1\neq l_2\ \ \ }} (I_{l_1}\cap I_{l_2})\}. 
\end{equation}
\textit{Step 3}: Since there is no two packets in the same subset in Step 1, there is no XORing opportunity in Step 3. Therefore, according to \eqref{eq:19}, we have
\begin{align}
     \beta_{CAPM}&=\sum_{p\in [q]} \max_{(i,j) \in K_p} H(S_{K_p} \mid A_{i}^{j}) \nonumber \\
     &=\max_{(i,j)} H(S_{V_{1,\dots,k}^{1}\cup V_{1,\dots,k}^{\prime}} \mid A_{i}^{j}) \nonumber \\
     &=\frac{k(k-1)}{2}-(k-2) \nonumber \\
     &=\frac{k(k-3)}{2}+2 \nonumber.
\end{align}
\end{IEEEproof}
\section{Conclusion}\label{sec:6}
In this paper, a new coding scheme, referred to as independent user partition multicast (IUPM) was proposed for the groupcast index coding (GIC) problem. The proposed coding scheme is based on user partitioning and removing the redundancies in the solution by eliminating the coded  packets  for  the  subproblems that are linearly dependent.  In other words, to further improve the performance, after partitioning the users into subsets, the proposed IUPM scheme uses MDS code to satisfy the users in each subset, however it just keeps those coded packets that are linearly independent. The reason is that the coded packets for some subsets can be achieved through other transmissions. Furthermore, a rigorous comparison between the PPM and UPM schemes was provided. It was proved that the UPM scheme includes the PPM scheme as a special case. Moreover, based on the user partitioning idea, a new polynomial-time heuristic algorithm for general GIC problems was proposed, which can be considered as a modified version of the CAPM scheme in \cite{Unal2016}.
\bibliographystyle{IEEEtran}
	\bibliography{References}

\end{document}